\documentclass[reprint]{revtex4}%
\usepackage{amsfonts}
\usepackage{amsmath}
\usepackage{amssymb}
\usepackage{charter}
\usepackage{graphicx}
\usepackage[colorlinks,citecolor=green,linkcolor=blue,anchorcolor=blue]{hyperref}
\usepackage[perpage]{footmisc}
\providecommand{\U}[1]{\protect \rule{.1in}{.1in}}
\begin{document}
\title{The role of the non-Gaussianity plays in the enhancement of the fidelity in continuous variable quantum teleportation}

\author{Ke-Xia Jiang\footnote{kexiajiang@126.com,kexiajiang@yeah.net}}

\affiliation{Department of Physics, Engineering University of
CAPF, Xi'an 710086, P. R. China}

\begin{abstract}
We investigated the role of non-Gaussianity (nG) plays in the enhancement of the fidelity in continuous-variable quantum teleportation of ideal Braunstein and Kimble (BK) protocol for coherent states, theoretically. The de-Gaussification procedure is realized through subtracting photons on the two-mode squeezed vacuum state (TMSVs). We find that the high fidelity always refers to a symmetrical arrangement of photon subtractions on the different modes of the TMSVs. The non-Gaussian resources demonstrate commendable superiorities compare with the Gaussian resources only for symmetrical arrangements of photon subtractions, however, the asymmetrical arrangements do not. When the total number of photon subtractions be a constant, we find that the optimal nG procedure prefers the most asymmetrical arrangement of subtractions. This characteristic is not consistent with the result that the highest fidelity refers to a symmetrical case. Under the same squeezing parameter, a higher nG might not always lead to a higher fidelity.
\end{abstract}
\maketitle

\section{Introduction}
In recent years, quantum information and communication (QIC) technologies of continuous variables have been discussed vastly in literatures focused on the entangled non-Gaussian optical states as communication resources~\cite{Opatrny2000,Cochrane2002,Olivares2003,Olivares2006,Dell'Anno2007,Dell'Anno2008,Dell'Anno2010a,Dell'Anno2010b,YangYang2009,Sabapathy2011}. This is mainly because their strongly nonclassical properties, such as the enhanced entanglements and clear negativity of the Wigner functions, which are important for the improvement of the efficiency in the implementation of QIC~\cite{Kim2005,LiHR2011,Chen2008,Kitagawa2006,L.-Y. Hu2010a,L.-Y. Hu2010b,Navarrete2012,Biswas2007,Zhang2010,Bartley2012,Mista2006}. In turn, to circumvent the restriction of no-go theorem~\cite{Eisert2002,Fiurasek2002,Giedke2002,Eisert2003} for entanglement distillation of Gaussian states, it has report that using non-Gaussian states and operations are feasible protocols~\cite{Ourjoumtsev2007,Takahashi2010}. It has also been shown that the non-Gaussian states are necessary to implement universal quantum computations~\cite{Lloyd1999,Menicucci2006,Gu2009}.

At present, non-Gaussian quantum states can be feasibly obtained through probabilistic operating procedures in laboratory~\cite{Zavatta2004,Parigi2007,Dakna1997,Wenger2004,Wakui2007,Ourjoumtsev2006,Gerrits2010}. A class of non-Gaussian quantum optical states can be realized by adding or subtracting an arbitrary number of photons on the TMSVs~\cite{Dakna1997,Wenger2004,Ourjoumtsev2006,Wakui2007,Kim2008,Gerrits2010}. However, due to the rapid decrease of the success probability, it becomes increasingly more challenging to construct large number of photon subtracted or added non-Gaussian states. Recently, Namekata et al.~\cite{Namekata2010} reported the experimental demonstration of a one- or two-photon subtracted squeezed state.

Theoretically, Dell' Anno et al.~\cite{Dell'Anno2007,Dell'Anno2008,Dell'Anno2010a,Dell'Anno2010b} investigated the performance of a class of two modes squeezed Bell-like entangled resources that take photon subtracted and photon added as particular instances in continuous-variable quantum teleportation of ideal and imperfect BK protocols~\cite{BK1998}. They found that the optimized non-Gaussian entangled resources outperformed their counterpart Gaussian resources in the implementation of QIC. However, the amount of entanglement of the communication resources are not the only factors for optimizing the teleportation fidelity. Though the same degree of entanglement, different procedures for de-Gaussification, generally, gave out different enhancements of the fidelity. Yang et al.~\cite{YangYang2009}. investigated the relations of the fidelity with the Einstein-Podolsky-Rosen correlations. And they found that in weak squeezing regions of certain resource it can carry out a high fidelity.

In this paper, we generally investigate the performance of a class of two mode non-Gaussian entangled resources Eq.~\eqref{TPSSV} in continuous-variable quantum teleportation of ideal BK protocol for coherent states Eq.~\eqref{characteristicCS}. The non-Gaussian entangled resources are obtained by performing local photon subtracted operations $a$ and $b$ on the two modes of TMSVs, respectively. In some sense, we generalize the non-Gaussian entangled communication resources of the work in Refs.~\cite{Cochrane2002,Olivares2006}. However, in our present work we analysis the role of nG plays in the enhancement of the fidelity, emphatically. Effective arrangements of photon subtraction operations for the improvement of teleportation fidelity are discussed.

This work is motivated by the following reasons. Firstly, in principle, the nG is not directly related to the nonclassical character of a quantum state. It has shown that the nG and entanglement of photon subtracted states have radically different behaviors~\cite{Navarrete2012}. Although the fidelity of teleportation can be improved by various de-Gaussification procedures~\cite{Cochrane2002,Olivares2006,Dell'Anno2007,Dell'Anno2008,Dell'Anno2010a,Dell'Anno2010b,YangYang2009}, it is not clear what role the nG plays in teleportation protocols, and whether a higher nG always leads to a higher fidelity. We adopt the measure proposed by Genoni et al.~\cite{Genoni2008,Genoni2010} to investigate the role of nG in this issue. Secondly, effective arrangements for de-Gaussification operations are considered. How the fidelity of teleportation can be optimized through properly arranging several limited de-Gaussification operations? In particular, we analyze the asymmetrical and also the symmetrical arrangements of the operations on different modes of the TMSVs where the total number of photon subtractions is constant. Lastly, considering photon subtraction technologies have been used for various aims in experiments, and further theoretical analysis have shown that photon subtractions are more efficient from the energy-cost point of view for increasing entanglements of communication resources~\cite{Navarrete2012}, only photon subtraction operations are included in our present work. However, the whole processing is also suitable for testing photon additions and other schemes for generating non-Gaussian states using the TMSVs~\cite{Parigi2007,YangYang2009}.

This paper is organized as follows: In Section~\ref{sec02}, for self-content, we introduce the theoretical expressions of the photon-subtracted TMSVs, and the photon subtraction techniques in laboratories, briefly. In Section~\ref{sec03}, using the non-Gaussian entangled resources Eq.~\eqref{TPSSV}, we generally investigate the teleportation fidelity of ideal BK protocol for coherent states~\eqref{characteristicCS}. The relationships among the teleportation fidelity with the entanglement and the nG  are discussed in  Section~\ref{sec04}. The final section, Section~\ref{sec05}, is devoted to conclusions and discussions.
\section{The arbitrary number of Photon-subtracted TMSVs}\label{sec02}
Theoretically, the arbitrary number of photon-subtracted TMSVs can be obtained by repeatedly operating the photon annihilation operators on a TMSVs. The normalized form can be written as~\cite{Quesne2001,L.-Y. Hu2010a,L.-Y. Hu2010b}
\begin{equation}\label{TPSSV}
|\psi\rangle =\frac{1} {\sqrt{N_{r,m,n}}} a^{m}b^{n}S(r)|00\rangle,
\end{equation}
where $m$, $n$ are the number of arbitrary subtracted photons on the two different modes (for Alice and Bob). The square of the normalization
constant of photon-subtracted TMSVs takes
\begin{equation}\label{normalization}
  {N_{r,m,n}}=
  \left\{
   \begin{aligned}
  &m!n!\sinh^{2n}r P^{(n-m,0)}_m(\cosh2r) & \hspace{10mm} &(m\leq n)  \\
  &m!n!\sinh^{2m}r P^{(m-n,0)}_n(\cosh2r) & \hspace{10mm} &(m\geq n)  \\
   \end{aligned}
  \right.,
\end{equation}
and $P_{m}^{(\alpha,\beta)}(x)$ is the Jacobi polynomial which defined as
$$
P_{m}^{(\alpha,\beta)}(x)=\left(  \frac{x-1}{2}\right)  ^{m}\sum_{k=0}%
^{m}\left(
\begin{array}
[c]{c}%
m+\alpha \\
k
\end{array}
\right)  \left(
\begin{array}
[c]{c}%
m+\beta \\
m-k
\end{array}
\right)  \left(  \frac{x+1}{x-1}\right)^{k}.$$ $S(r)=\exp
[r(a^{\dag} b^{\dag}-ab)] $ is the two-mode squeezing operator
with $r$ being a real squeezing parameter, and the Bose
annihilation and creation operators satisfy
$[a,a^\dag]=[b,b^\dag]=1$. When $m=n=0$, the normalization constants
reads $1/\sqrt{N_{r,0,0}}=1$, and the states reduces to the familiar original Gaussian TMSVs
\cite{Reid1988,Reid1989,Walls1994}
\begin{equation}\label{002}
|\textrm{TMSV}\rangle=S(r)|00\rangle=\sqrt{1-\lambda^2}\sum_{n=0}^{\infty}\lambda^n|nn\rangle,
\end{equation}
where $\lambda=\tanh r$.

The photon subtraction process on TMSVs was first proposed by Opatrn\'{y} et al.~\cite{Opatrny2000}, theoretically. In laboratories, photon subtraction can be done by splitting a small fraction of the squeezed vacuum for photon counting~\cite{Neergaard-Nielsen2006,Ourjoumtsev2006,Wakui2007}, which are guided into a photodetector, and by selecting the transmitted state conditioned on the detection of photons which can be signed as ``clicks" in an on/off photodetector, e.g. the avalanche photodiodes. Although subtracting large number of photons leads to complex states, experiments get more challenging due to the rapid decrease of the success probability\cite{Takahashi2008}.

\section{The teleportation fidelity using the non-Gaussian entangled resources obtained by performing photon subtracted operations}\label{sec03}
\subsection{The teleportation fidelity}\label{sec031}
The success probability to teleport a pure quantum state can be described through the fidelity of teleportation, $F=\text{tr}(\rho_{\text{in}}\rho_{\text{out}})$, which as a measure of how close between the initial input state and the final (mixed) output quantum state. In the formalism of the characteristic functions of continuous variables, the fidelity can be written as~\cite{Chizhov2002}
\begin{equation}\label{fidelity}
    F=\frac{1}{\pi}\int d^2\alpha \chi_{\text{in}}(\alpha)\chi_{\text{out}}(-\alpha),
\end{equation}
where $d^2\alpha\equiv d \text{Re}(\alpha) d \text{Im}(\alpha) $. Using the Weyl displacement operator $D(\alpha)=\exp (\alpha a^\dag-\alpha^* a)$, the characteristic function of the single-mode input coherent state $|\mu\rangle= D (\mu)|0\rangle$ reads
 \begin{equation}\label{characteristicCS}
    \chi_{\text{in}}(\alpha)=\langle \mu|D(\alpha)|\mu\rangle=\exp \left[-\frac{|\alpha |^2}{2}+2 i \text{Im}(\alpha \mu ^*)\right].
\end{equation}
For the output teleported state, the characteristic function has the factorized form~\cite{Marian2006}
\begin{equation}\label{factorizedform}
    \chi_{\text{out}}(\alpha)=\chi_{\text{in}}(\alpha)\chi_{12}(\alpha^{*},\alpha),
\end{equation}
where $\chi_{12}(\alpha,\beta)$ is the symmetrically ordered characteristic function of the entangled resource. For the non-Gaussian entangled states Eq.~\eqref{TPSSV}, Eq.~\eqref{factorizedform} reads
\begin{equation}\label{characteristicnG01}
    \chi_{12}(\alpha,\beta)=\langle \psi|D_1(\alpha)D_2(\beta)|\psi\rangle,
\end{equation}
where $D_1(\alpha)=\exp (\alpha a^\dag-\alpha^* a)$ and $D_2(\beta)=\exp (\beta b^\dag-\beta^* b)$ correspond to the two modes of the entangled states, respectively. Using the relations of operators
\begin{align}
    e^{-\frac{|\alpha|^2}{2}} \Lambda^{m}(\alpha)D_1(\alpha)e^{\frac{|\alpha|^2}{2}}&=a^{\dag m}D_1(\alpha)a^{m},\label{relations01}\\
    e^{-\frac{|\beta|^2}{2}}\Lambda^{n}(\beta)D_2(\beta)e^{\frac{|\beta|^2}{2}}&=b^{\dag n}D_2(\beta)b^{n}\label{relations02},
\end{align}
with $\Lambda^{m}(\alpha)=\left(-\frac{\partial}{\partial\alpha} \right)^m \left(\frac{\partial}{\partial\alpha^*} \right)^m $ and $\Lambda^{n}(\beta)=\left(-\frac{\partial}{\partial\beta} \right)^n \left(\frac{\partial}{\partial\ \beta^*} \right)^n $, Eq. \eqref{characteristicnG01} can be reduced to a simple form
\begin{equation}\label{characteristicnG02}
   \chi_{12}^{(m,n)}(\alpha,\beta)={N^{S}_{\lambda,m,n}}^{-1} e^{ -\frac{1}{2}(|\alpha|^2+|\beta|^2)}\Lambda^{m}(\alpha)\Lambda^{n}(\beta)\left[ \chi(\alpha,\beta)e^{ \frac{1}{2}(|\alpha|^2+|\beta|^2)}\right],
\end{equation}
where the characteristic function of the TMSVs Eq. \eqref{002} reads
\begin{equation}\label{characteristicTMSVs}
     \chi(\alpha,\beta)=\exp\left[ -\frac{1+\lambda^2}{2(1-\lambda^2)}\left(|\alpha|^2+|\beta|^2\right)+\frac{\lambda}{1-\lambda^2}(\alpha\beta+\alpha^*\beta^*)\right],
\end{equation}
and $(m,n)$ denotes the number of photon subtracted operations on different modes (1 and 2) of the TMSVs. According to the above expressions, for the arbitrary number $(m,n)$ of photon subtractions, the fidelity for teleporting a coherent state based on the non-Gaussian entangled resources can be obtained, which is a function of the squeezing parameter, viz. $F=F(\lambda)$. The fidelity is equal for the two modes (1 and 2) of photon subtracted operations and this can be represented as
\begin{equation}\label{appendix01}
    F=F^{(m,n)}=F^{(n,m)}.
\end{equation}

For special cases, benefit from the number expressions of the series expansions for $\lambda$, we find the results which are listed in the \ref{appendix}. However, because of the arbitrary order partial derivatives in Eq.~\eqref{characteristicnG02}, finding general expressions come across challenges.

\subsection{Effective arrangements of photon subtractions}\label{sec032}
Using the results in the \ref{appendix}, it is straightforward to discuss the effective arrangements of photon subtractions on the TMSVs for the improvement of teleportation fidelity.

\begin{figure}[htbp]\centering
\includegraphics[width=2.8in]{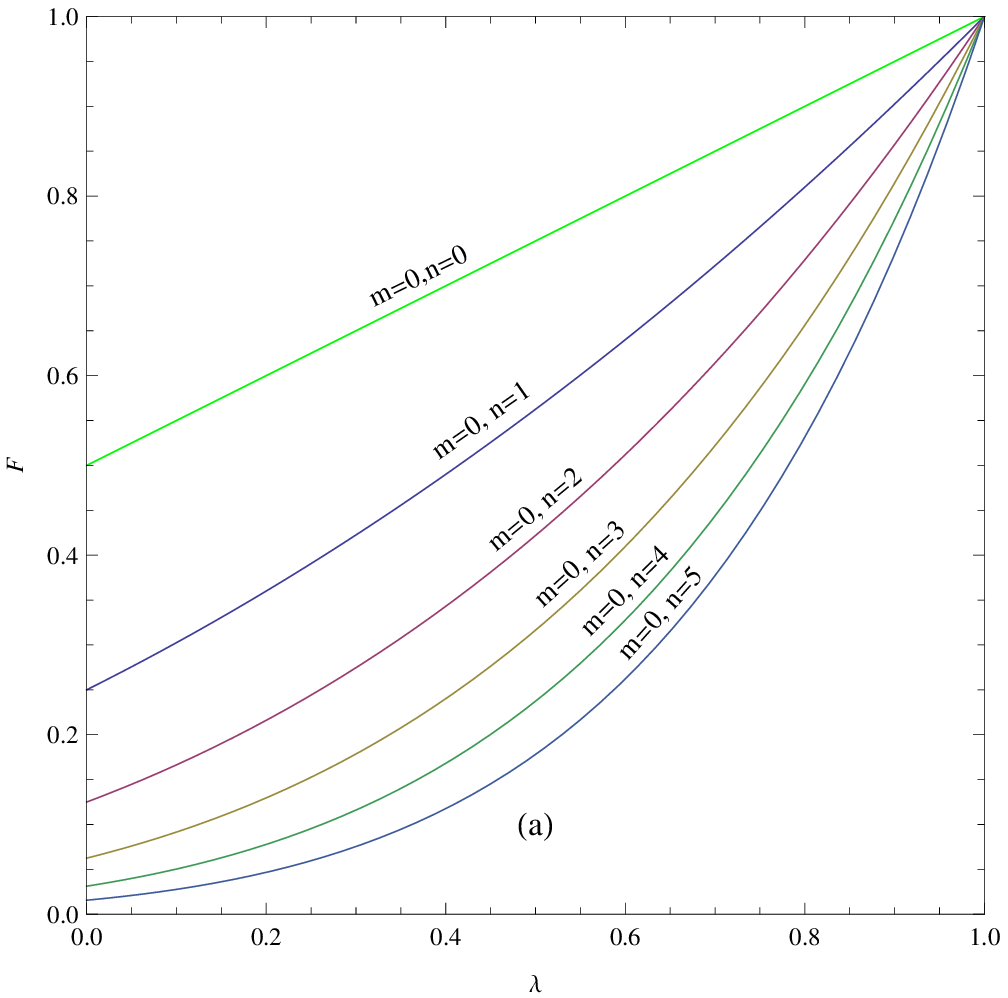}
\includegraphics[width=2.8in]{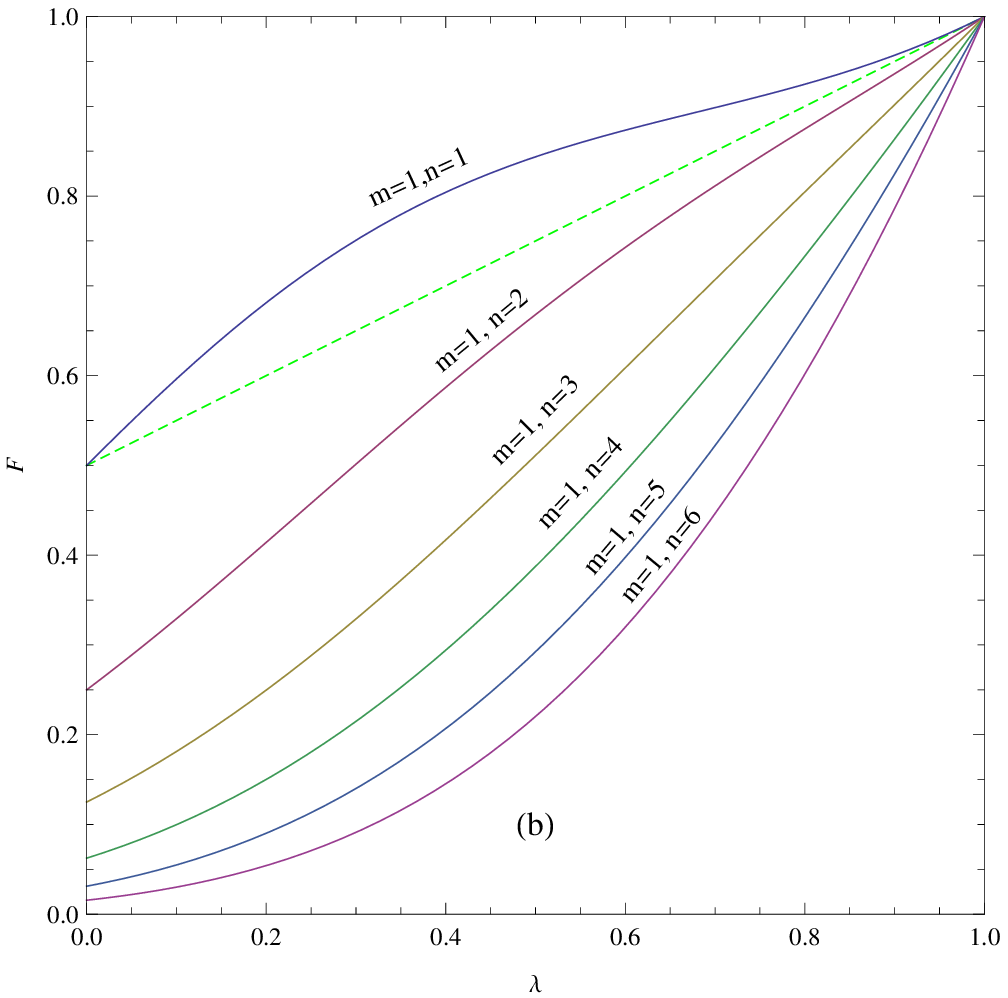}\\
\includegraphics[width=2.8in]{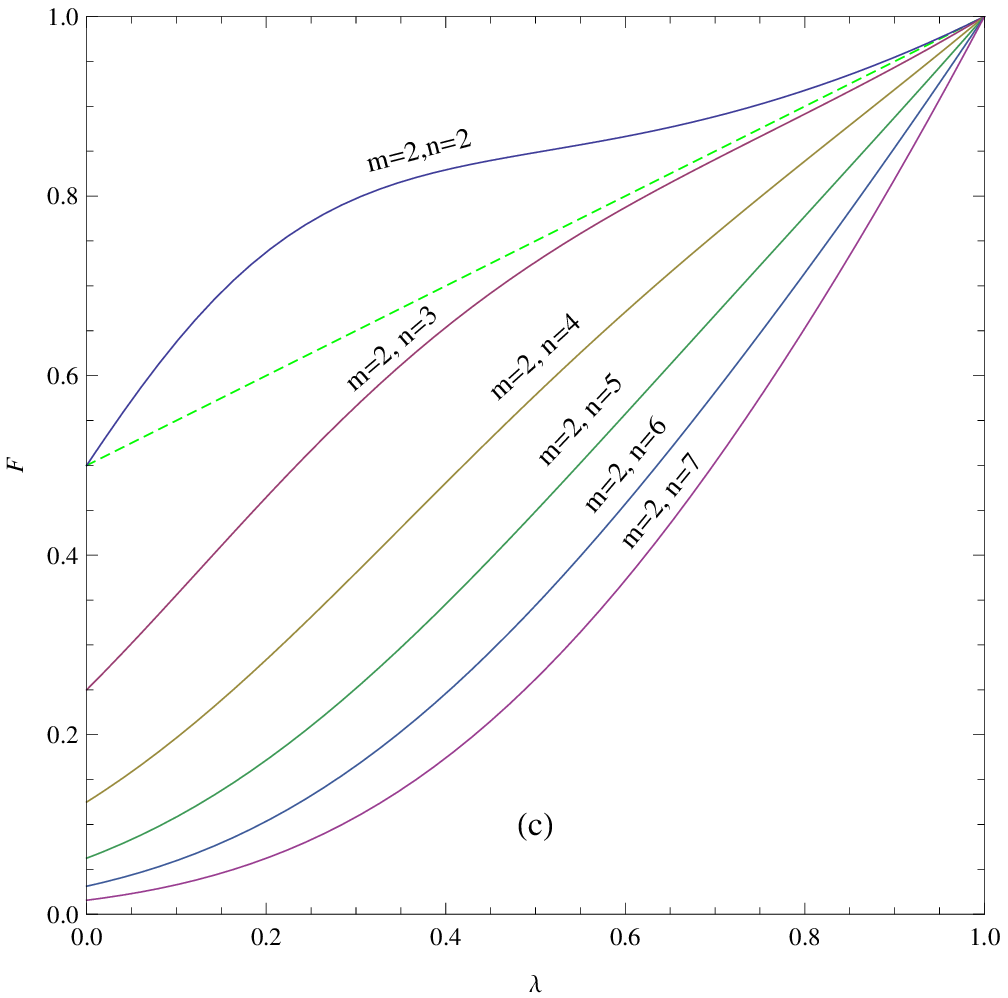}
\includegraphics[width=2.8in]{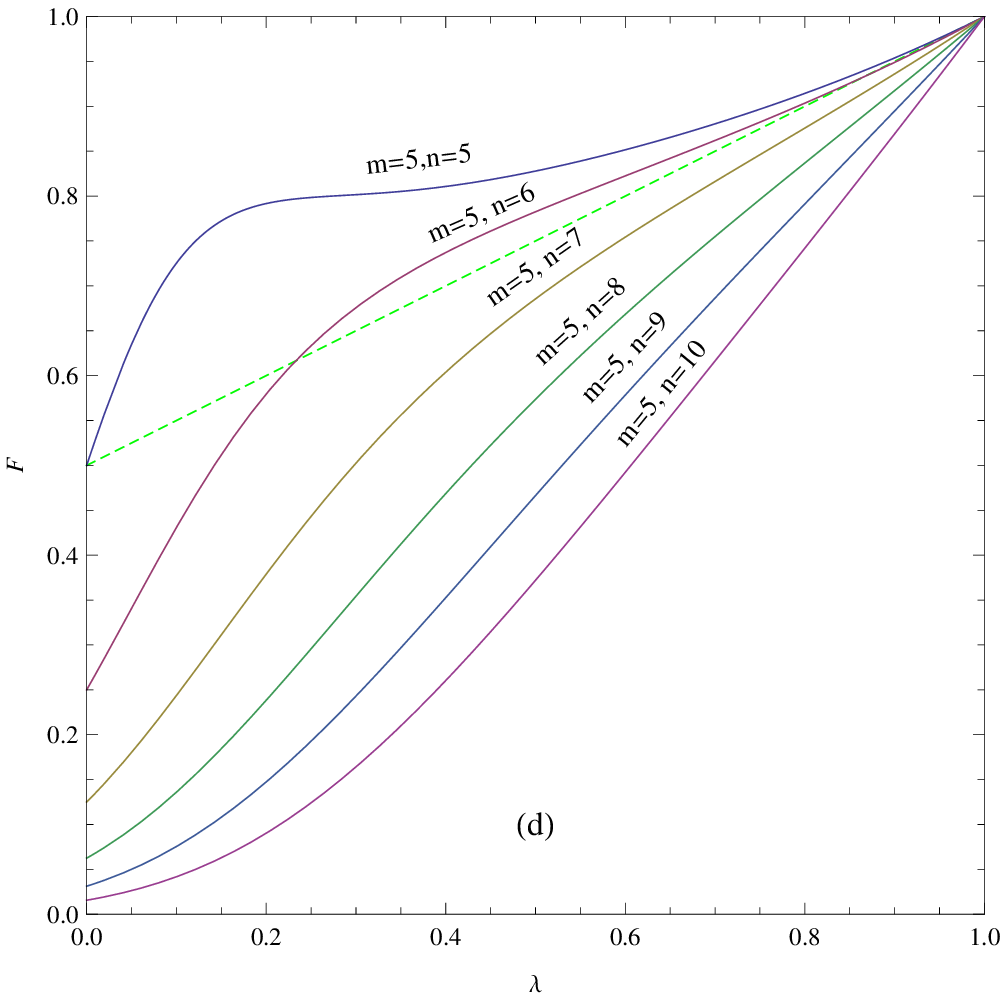}\caption{(Color online)
Fidelities as functions of the squeezing parameter with $n\geq m$ and (a) $m=0$; (b) $m=1$; (c) $m=2$; (d) $m=5$. For the same squeezing parameter, higher fidelities always refer to symmetrical arrangements of photon subtractions on the different modes of the TMSVs. The fidelity for the Gaussian resource ($m=n=0$) is attached in dashed green (dashed) lines.
}\label{Fig1}
\end{figure}
\begin{figure}[htbp]\centering
\includegraphics[width=2.8in]{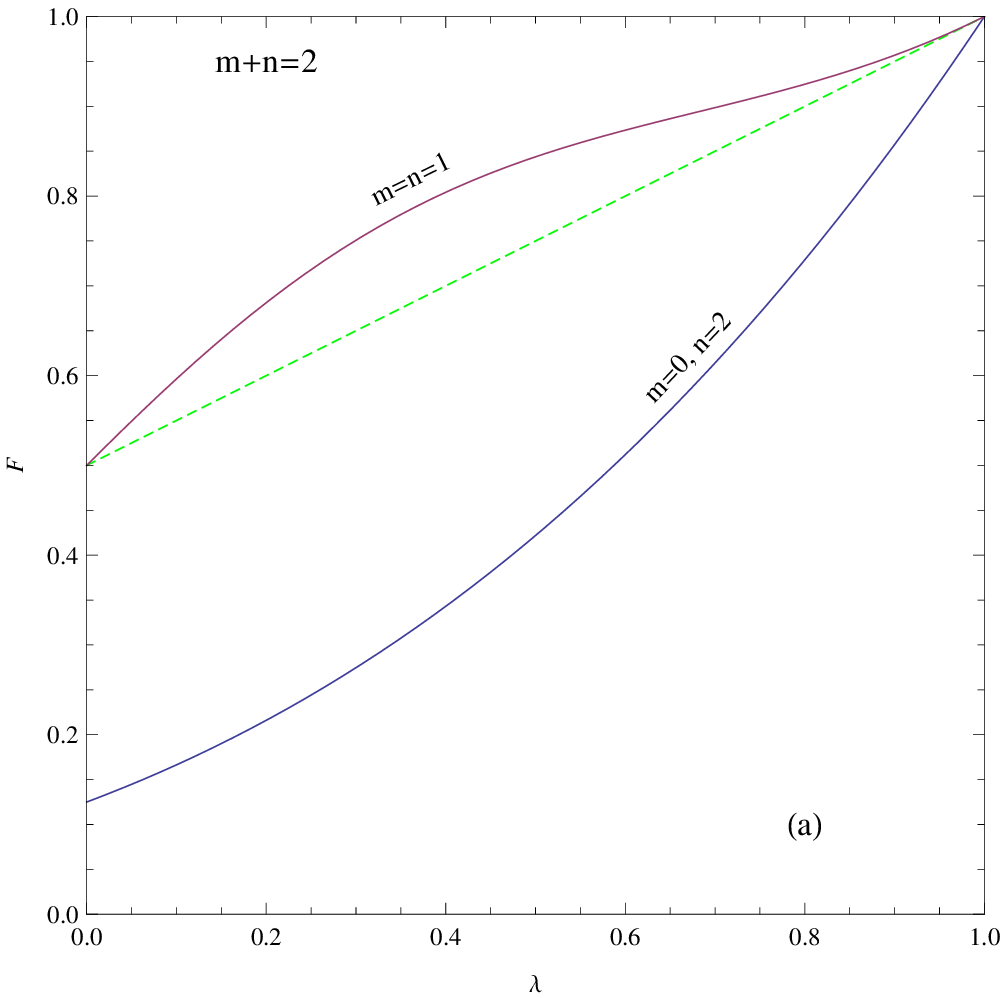}
\includegraphics[width=2.8in]{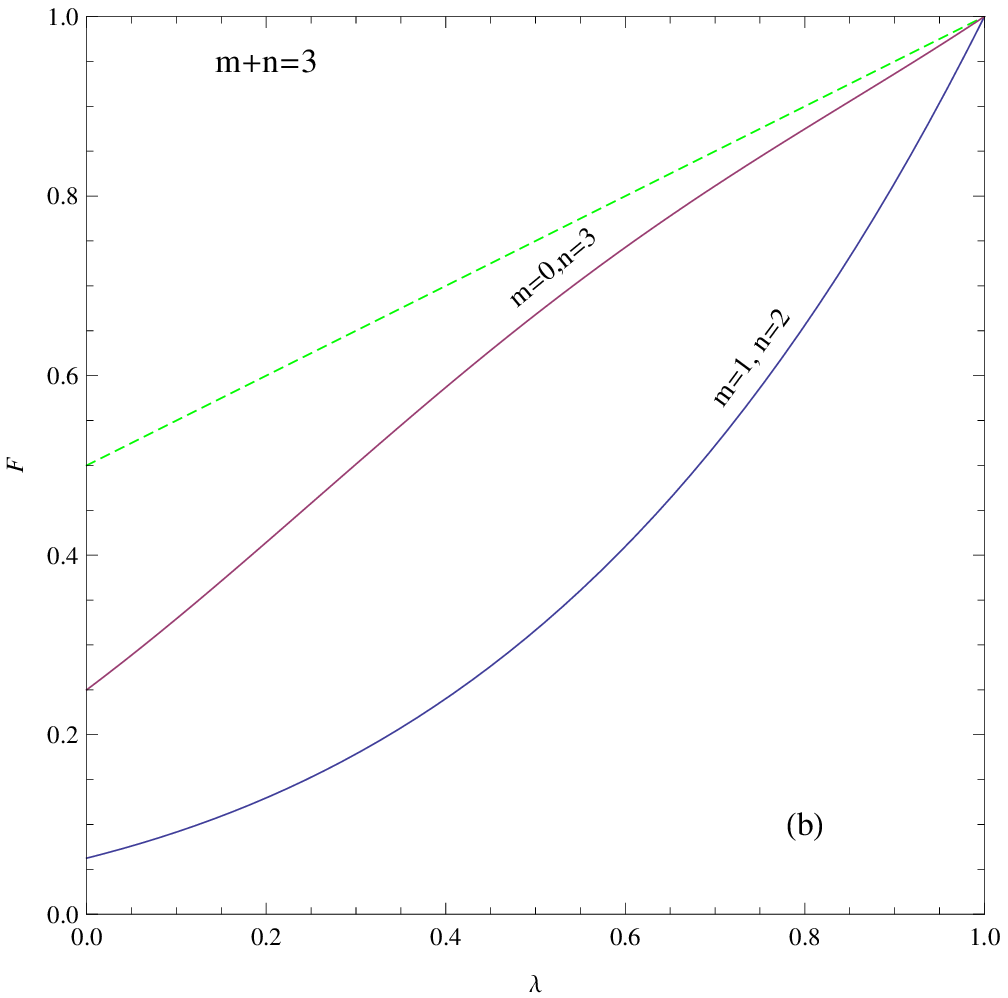}\\
\includegraphics[width=2.8in]{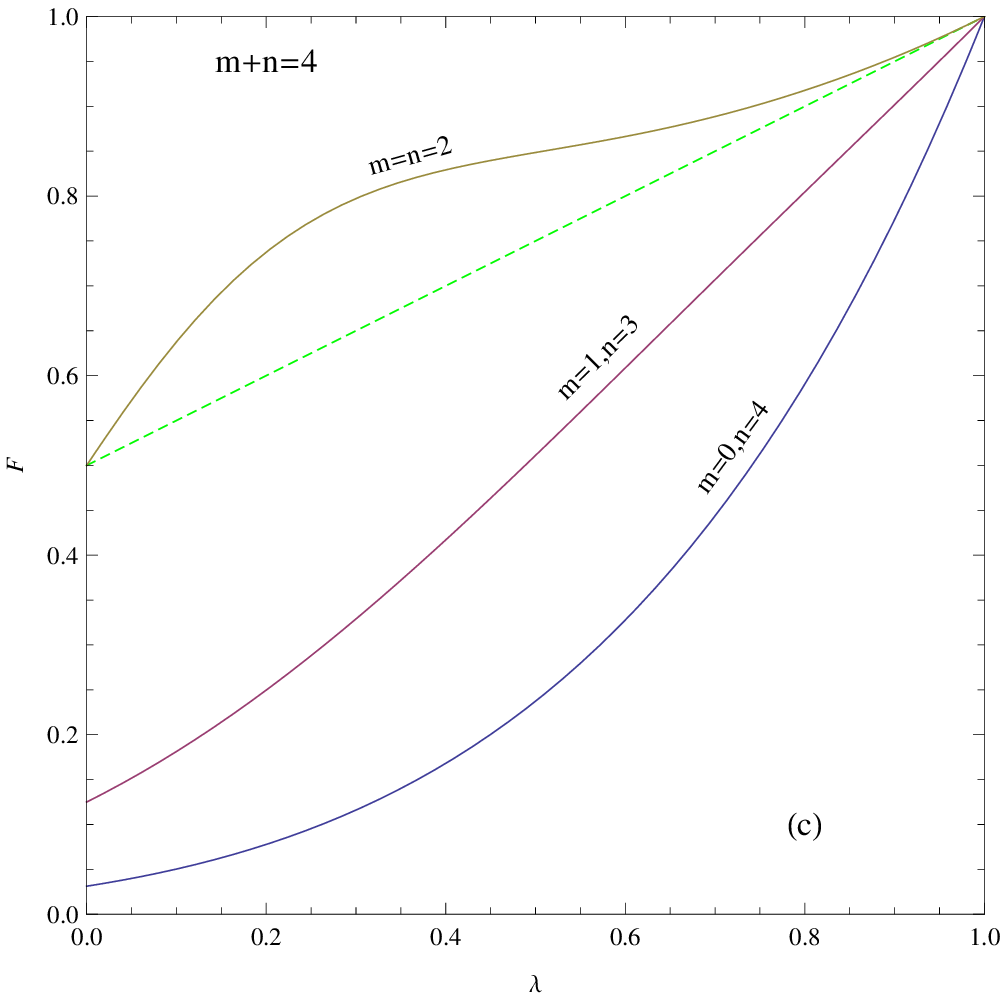}
\includegraphics[width=2.8in]{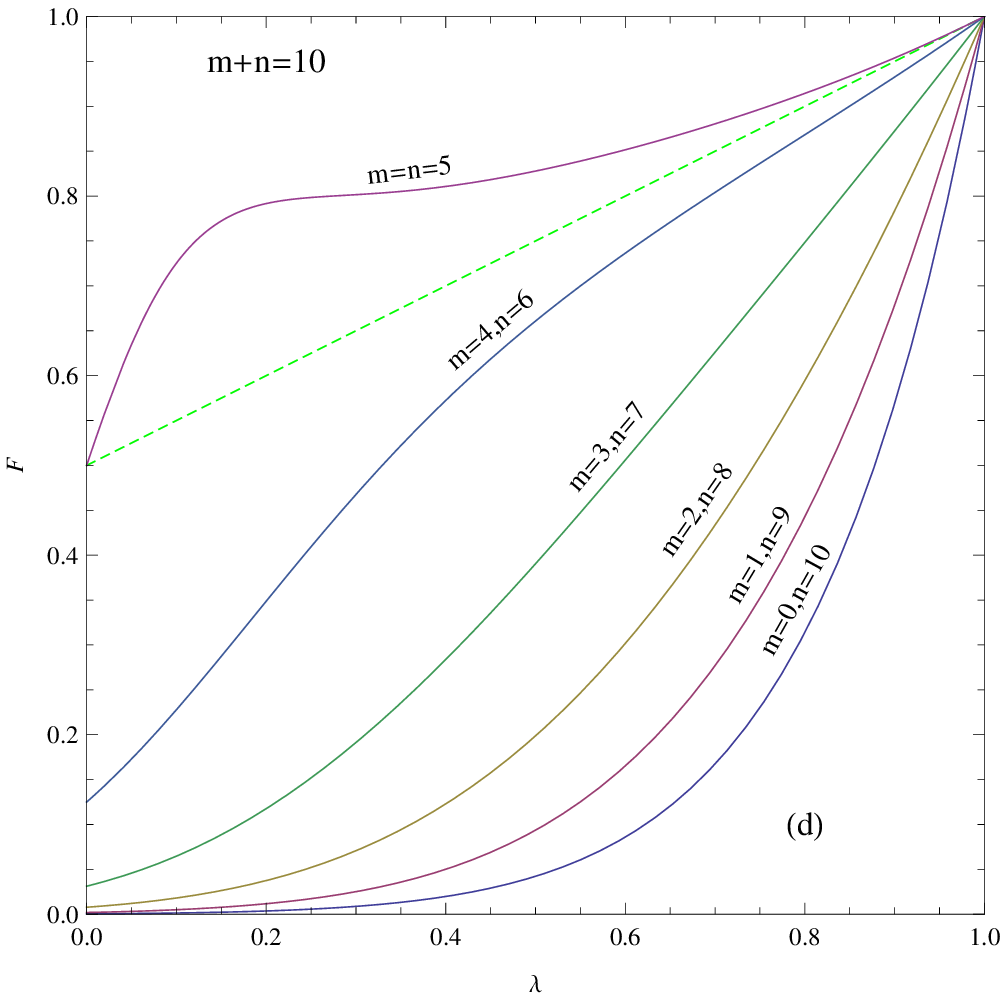}\caption{(Color online)
Fidelities as functions of the squeezing parameter with the total number of photon subtractions be a constant $C$: (a) $C=2$; (b) $C=3$; (c) $C=4$; and (d) $C=10$. And the fidelity for the Gaussian resource ($m=n=0$) is also attached (green dashed lines).
}\label{Fig2}
\end{figure}
(i) Firstly, considering the equality of the two modes, only cases $n\geq m$ (with $m=0,1,2,\ldots$) are enough. We plot the series in Fig.~\ref{Fig1}. It is shown that, for the same squeezing parameter, non-symmetrical arrangements lead to lower fidelities, while higher fidelities always prefer to symmetrical arrangements of photon subtractions on the two modes of the TMSVs. For small squeezing parameters, non-symmetrical arrangements even could not guarantee that the fidelities over the classical limit $1/2$. A significant result is that some non-symmetrical arrangements does not exhibit superior performances compare with the Gaussian resource ($m=n=0$) which is attached as green dashed lines (see Fig.~\ref{Fig1}(b)-(d) and Fig.~\ref{Fig2}).

\begin{figure}[htbp]\centering
\includegraphics[width=2.8in]{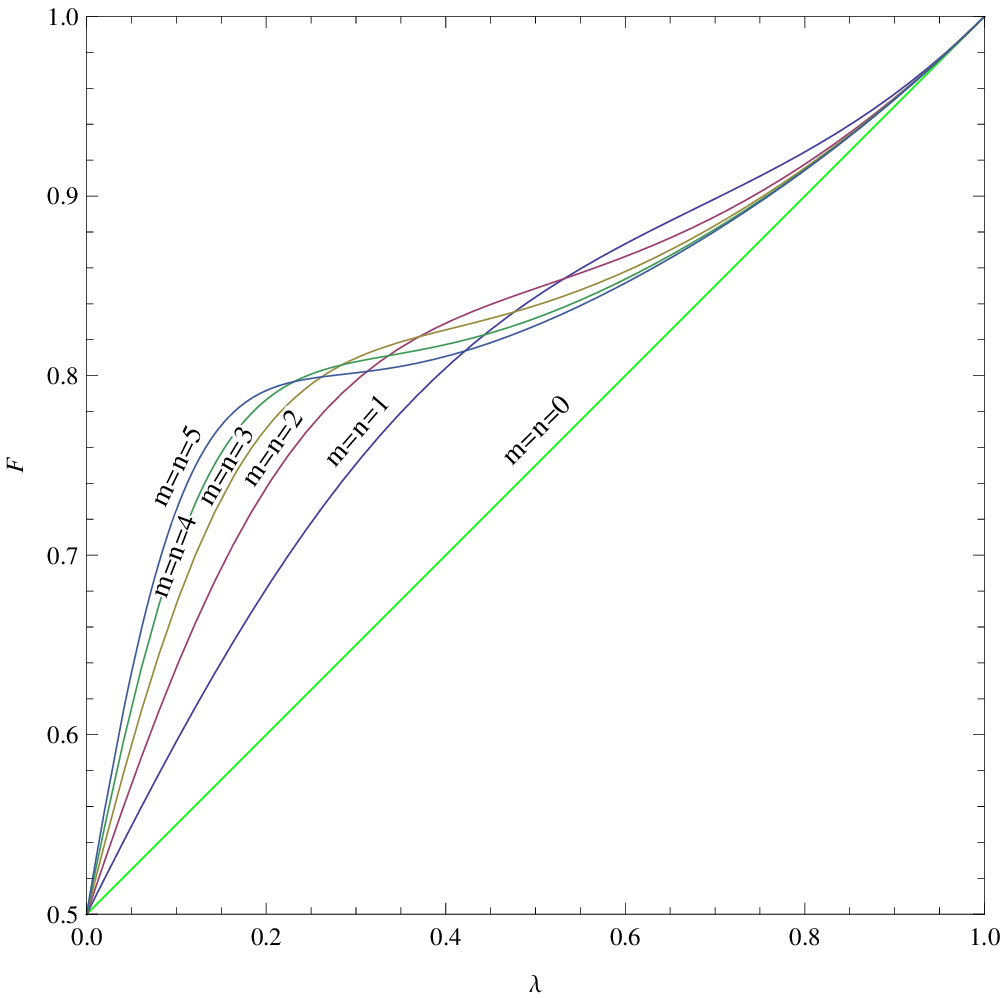}
\caption{(Color online)
Fidelities as functions of the squeezing parameter for symmetrical arrangements of photon subtractions. Although fidelities are not increased monotonically, they are being improved entirely compare with the Gaussian resources ($m=n=0$).
}\label{Fig3}
\end{figure}

(ii) Further, letting the total number of photon subtractions be a constant $C$ and from the curves in Fig.~\ref{Fig2}, it can be shown that high fidelities also prefer to symmetrical arrangements. We plot the series of symmetrical cases in Fig.~\ref{Fig3}. And it is shown that the fidelities are improved entirely, where the non-Gaussian resources demonstrate commendable superiorities compare with the Gaussian resources. However, fidelities are not monotonically increased with the number of operations.

These properties are very similar with the characteristics of entanglement in Ref.~\cite{Navarrete2012}, where it has shown that the highest entanglement is achieved when these operations are equally shared between the two modes.
\section{The teleportation fidelity with the non-Gaussianity of resources}\label{sec04}

In this section, we investigate the role of nG plays in teleportation protocols, where the measure proposed by Genoni et al. \cite{Genoni2008,Genoni2010} is adopted.

\begin{figure}[htbp]\centering
\includegraphics[width=2.8in]{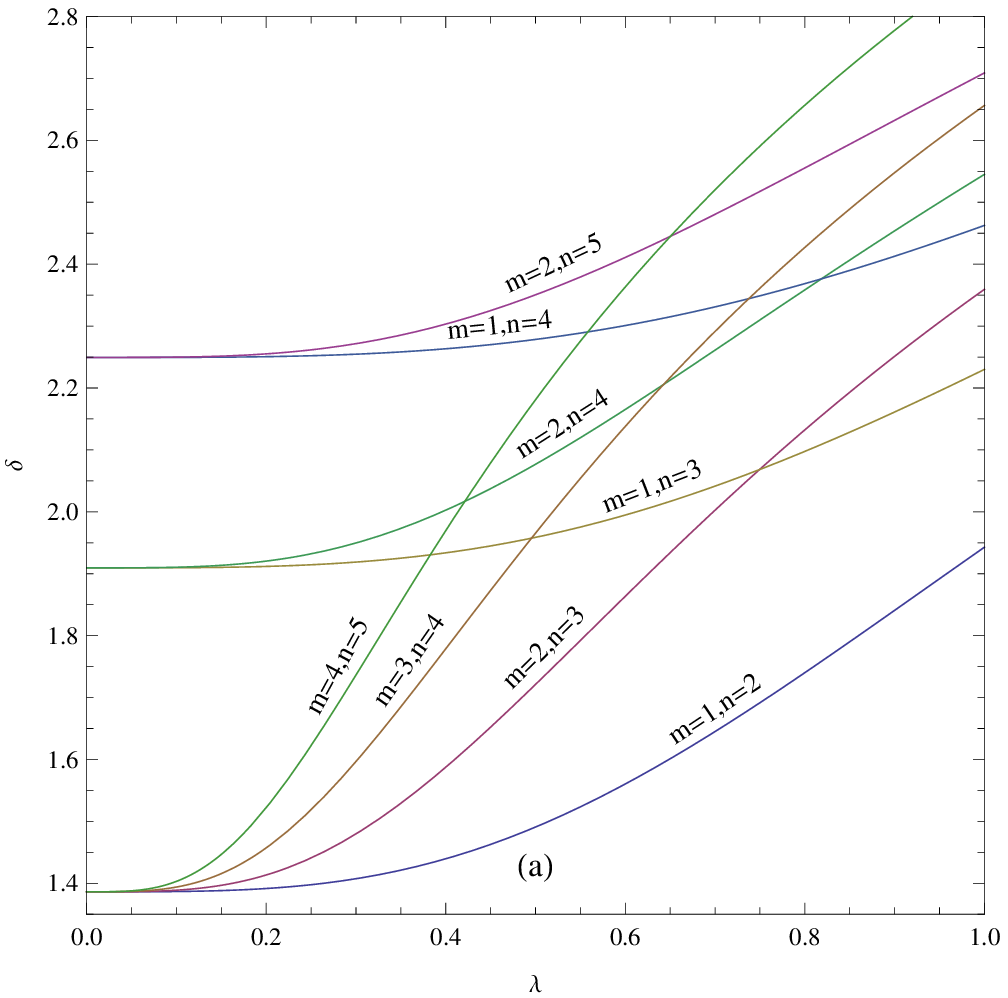}
\includegraphics[width=2.8in]{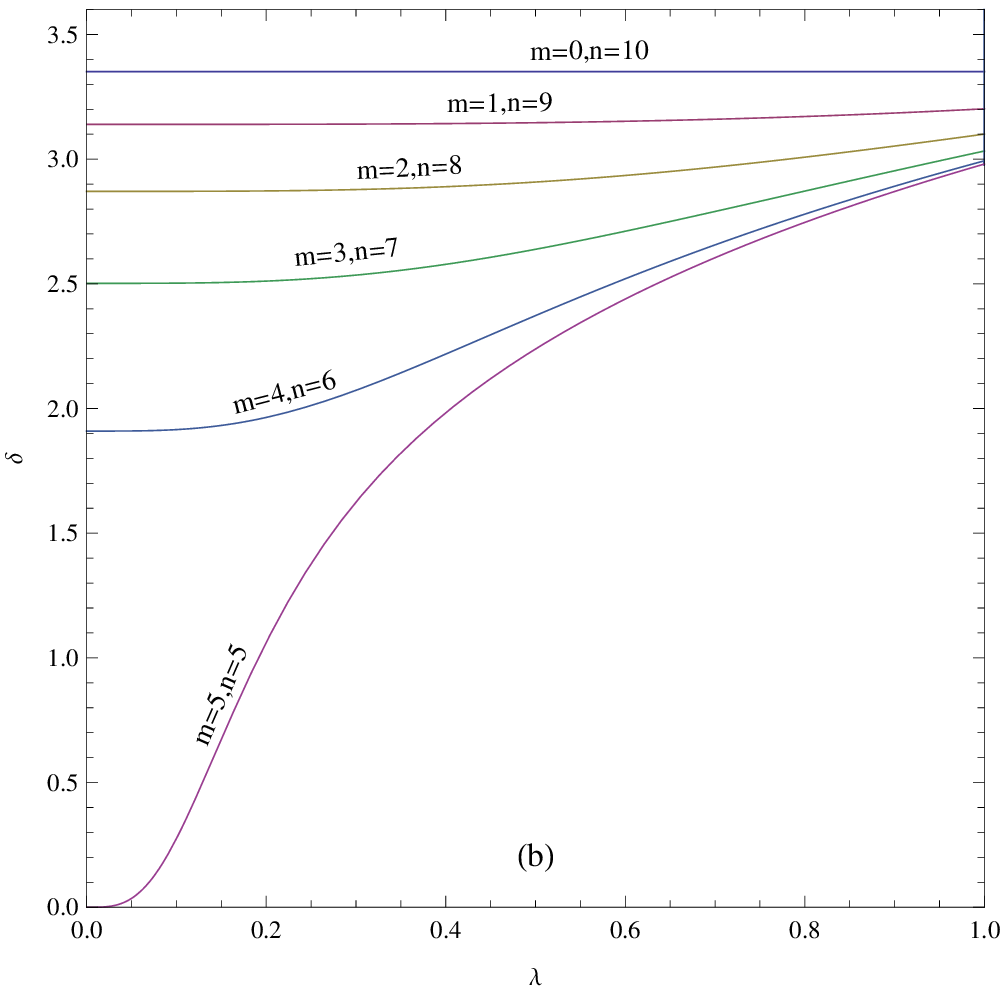}\\
\includegraphics[width=2.8in]{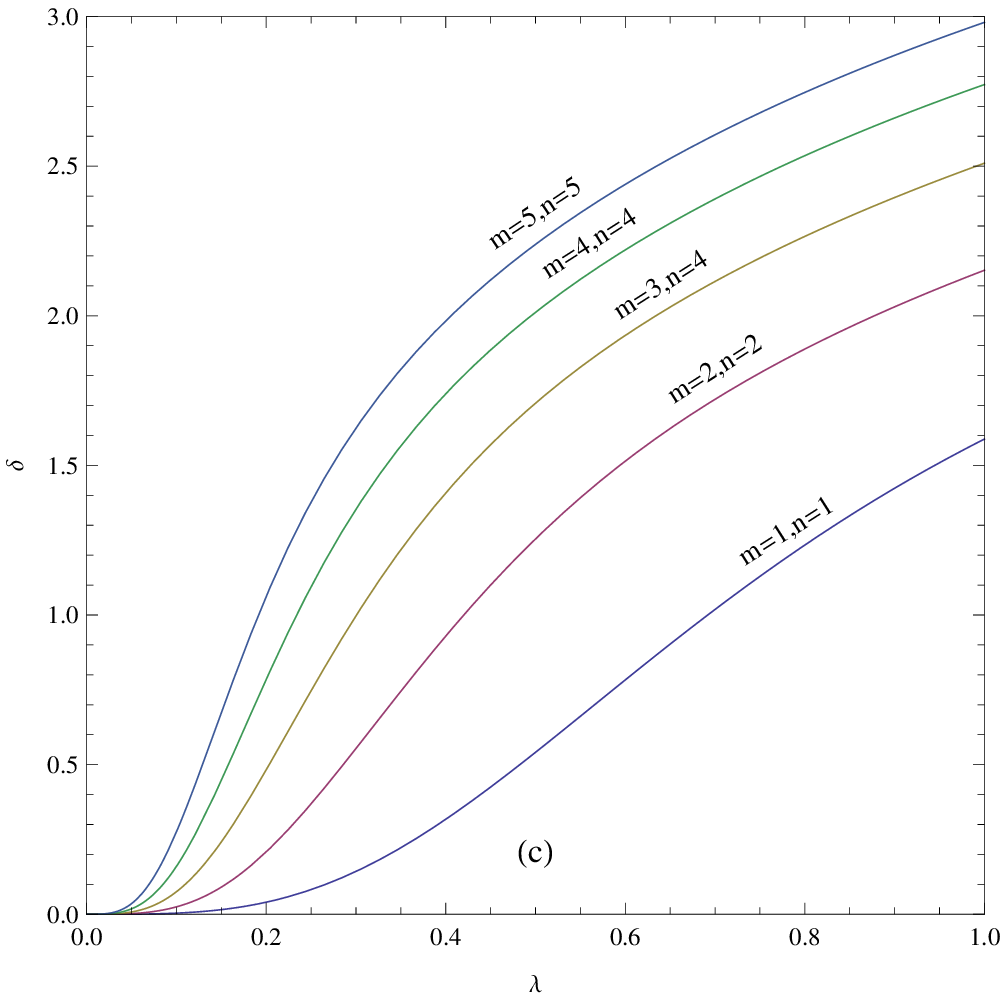}
\includegraphics[width=2.8in]{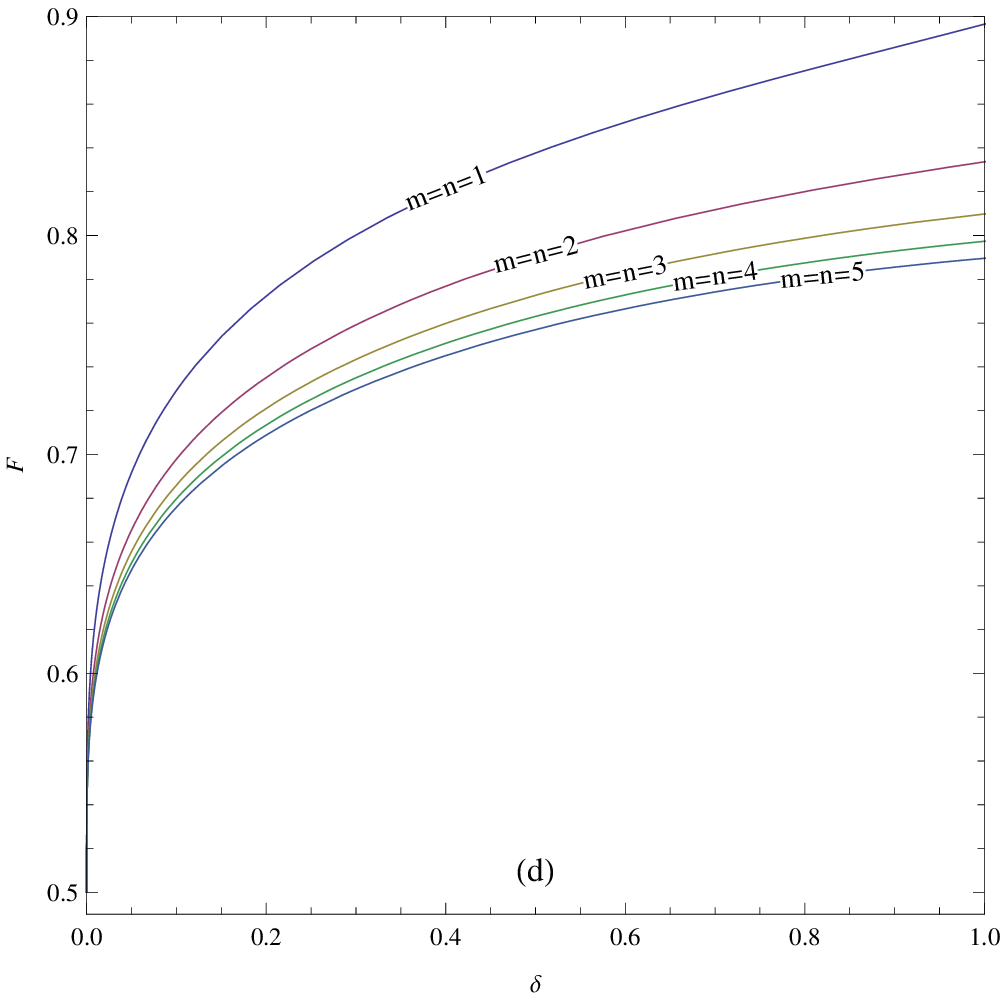}
\caption{(Color online)
The nG as the functions of the squeezing parameter in figures (a)-(c). (a) When $n-m=\text{Constant}$, the series have the same nG as the squeezing parameter comes to zero. (b) More asymmetrical arrangements of the operations on the two modes give more nG. (c) For symmetrical arrangements, a larger number operations always represent a higher nG. Figure (d) shows the relations of the fidelity with the nG for symmetrical cases, where $m=n=1$ represents the most rapidly increasing behavior.
}\label{Fig4}
\end{figure}
For a general state $\rho$, the degree of nG is quantified as the difference of the von Neumann entropy between the state and its reference Gaussian state
\begin{equation}\label{nGdef}
    \delta [\rho]=S(\rho||\tau)=\text{Tr}[\rho \text{ln}\rho]-\text{Tr}[\rho \text{ln}\tau]=S(\tau)-S(\rho).
\end{equation}
The reference Gaussian state $\tau$ has the same first- and second-order moments of canonical quadrature variables with the state $\rho$. The two moments are the vector of mean values $X[\tau]=X[\rho]$ and the covariance matrix (CM) $\sigma[\tau]=\sigma[\rho]$, respectively, which defined as
\begin{align}
    X&:=\langle R\rangle,\label{1thmoment}\\
    \sigma&:=\frac{1}{2}\langle\{RR^T+(R^TR)^T\}\rangle-\langle R\rangle \langle R^T\rangle,\label{2thmoment}
\end{align}
where the vector operator $R:=(x_1,p_1,\cdots, x_n,p_n)^T$. For the two-mode states, the formula can be reduced to
\begin{equation}\label{nGtwomode}
    \delta [\rho]=h(d_{-})+h(d_{+})-S(\rho),
\end{equation}
and the function $h(x)$ is defined as
\begin{equation}\label{defhx}
    h(x)=(x+\frac{1}{2})\text{ln} (x+\frac{1}{2}) -(x-\frac{1}{2})\text{ln} (x-\frac{1}{2}).
\end{equation}
Here $d_{\pm}$ are the symplectic eigenvalues of the two-mode CM
\begin{equation}\label{twomodeCM}
\sigma=
\left(
\begin{array}{c|c}
A  & C\\
\hline
C^T & B
\end{array}
\right),
\end{equation}
with $A$, $B$, and $C$ are $2\times 2$ real matrices. In the standard from, matrices $A$ and $B$ are proportional to the identity and $C$ is diagonal. Using the four local symplectic invariants $I_1 \equiv \det(A)$, $I_2 \equiv \det(B)$, $I_3 \equiv \det(C)$, and $I_4 \equiv \det(\sigma)$, the symplectic eigenvalues can be expressed as
\begin{equation}\label{symplecticeigenvalues}
    d_{\pm}=\sqrt{\frac{\triangle(\sigma)\pm \sqrt{\triangle(\sigma)^2-4I_4}}{2}}
\end{equation}
with $d_{+}\geq d_{-}$ and $\triangle(\sigma) =I_1+I_2+2I_3$. Actually, for pure stated $\rho$ since $S(\rho)\equiv 0$, the nG is always equal with the von Neumann entropy of the corresponding Gaussian states $\tau$. Using the definition of CM Eq.~\eqref{2thmoment}, one can obtain
\begin{align}\label{matriceA}
A&=\Big(\frac{1}{2}+\frac{N_{r,m+1,n}}{N_{r,m,n}}\Big)\openone,\\
B&=\Big(\frac{1}{2}+\frac{N_{r,m,n+1}}{N_{r,m,n}}\Big)\openone,\\
C&=\frac{1}{2}(n+1)\sinh(2r)\frac{P^{(n-m,1)}_m(\cosh2r)}{P^{(n-m,0)}_m(\cosh2r)}
\Big(
\begin{array}{cc}
1  & 0\\
0 & -1
\end{array}
\Big).
\end{align}
Without loss generality, we have assumed that $n\geq m$ in the above results.  Now, it is straightforward to give the nG for pure states Eq.~\eqref{TPSSV} according with the definition in Eq.~\eqref{nGtwomode}.

In Fig.~\ref{Fig4}(a)-(c), we plot the nG as the functions of the squeezing parameter. For cases with $m\cdot n\neq 0$, the nG is a monotonically increasing function. And for series with $n-m=\text{Constant}$, they have the same nG as the squeezing parameter comes to zero, see Fig.~\ref{Fig4}(a). The nG for series $n+m=10$ are ploted in Fig.~\ref{Fig4}(b). It is shown that the more asymmetrical arrangements of the operations on the two modes always prefer to higher nG, which is consistent with the findings in the literature~\cite{Navarrete2012}. When photon subtracted operations are only added on the one mode (e.g. the most asymmetrical arrangement $m=0$), the nG reaches a maximum $\delta[\rho]=-n \text{ln} (n)+(n+1) \text{ln}(n+1)$ (it becomes does not rely on the squeezing parameter). However, the optimal procedure for getting a highest nG prefers the most asymmetrical arrangement for the operations when $n+m=\text{Constant}$, is not consistent with the result that the highest fidelity always prefers to a symmetrical arrangement in Sec.~\ref{sec032}. For the series of symmetrical cases, it can be seen in Fig.~\ref{Fig4}(c) that a larger number operations always represent a higher nG. In Fig.~\ref{Fig4}(d), we plot the relations of the fidelity with the nG for symmetrical cases. Though all the fidelities are being raised due to the introduction of the nG, the case of $m=n=1$ represents the most rapidly increasing behavior.

\section{Conclusions and discussions} \label{sec05}
In this paper, we studied the effects on the fidelity of a de-Gaussification procedure on the entangled Gaussian communication resources in continuous-variable quantum teleportation of BK protocol for coherent states. The procedure is realized through subtracting photons on the TMSVs. We investigated the role of nG plays in the enhancement of the fidelity, emphatically.

We find that the high fidelity always prefers to a symmetrical arrangement of photon subtractions on the different modes of the TMSVs. When the total number of photon subtractions is a constant, the conclusion is still affirmative. Finding a general expression of the fidelity as a function of the squeezing parameter comes across challenges due to the arbitrary order partial derivatives. Using the mathematical induction method, we find the analytical results for some special cases which are listed in the \ref{appendix}.

A significant result is that non-Gaussian resources demonstrate commendable superiorities compare with the Gaussian resources only for symmetrical arrangements of photon subtractions. Non-symmetrical arrangements even could not guarantee that the fidelities over the classical limit $1/2$. They do not exhibit superior performances compare with the Gaussian resource.

The role of the nG plays in teleportation protocols is studied. When operations take $n+m=\text{Constant}$, the optimal de-Gaussification procedure prefers the most asymmetrical arrangement ($m=0$). This characteristic is not consistent with the result that the highest fidelity prefers to a symmetrical arrangement. In symmetrical cases, a larger number of operation represents a higher nG, however, the lowest case $m=n=1$ exhibits the most rapidly increasing behavior. Then at the same squeezing parameter, a higher nG might not always lead to a higher fidelity.

Considering the recent study in Ref.~\cite{Navarrete2012}, where the authors invested the quantum entanglement properties enhanced by photon addition and subtraction, in our work we find that the characteristics of the improved fidelities are more preferred to the enhanced quantum entanglement. While, the relations between the improved fidelities and the nG, which measured by the method of Genoni et al. \cite{Genoni2008,Genoni2010}, are subtle. In certain conditions, enhancing the nG is not always means improving the fidelity of the quantum teleportation.

The weakness of the present work is that we do not consider the rate of success of the probabilistic photon subtraction operations performing on the initial entangled TMSVs. In the very recent and interesting work~\cite{Bartley2012}, the authors analyzed different strategies of local photon subtraction from TMSVs in terms of entanglement gain and success probability. For the lossless limit case, where the loss of the fraction of photons is zero, they found the best strategy to produce the highest entanglement gain rate is subtracting a single photon only from one mode. Affirmatively, from our asymmetric cases in Fig.~\ref{Fig1}(a), one can find this strategy will do not work well since it even could not guarantee the fidelities over the classical limit $1/2$. For the lossy case, they found the best strategy is symmetric subtractions. Interestingly and fortunately, our result also shows that the high fidelity always prefers to a symmetrical arrangement of photon subtractions. However, a thoroughly analysis is interesting and needed, which is beyond the present work and will be discussed further in the future research directions.

\textbf{Acknowledgments}
The author thanks Ph.D San-Min Ke for very helpful discussions. This work was supported by the Basic Research Foundation of Engineering University of CAPF(WJY-201104),and partly supported by the National Natural Science Foundation of China under Grant Nos.61072034.
\appendix
\section{}\label{appendix}
Due to the arbitrary order partial derivatives in Eq. \eqref{characteristicnG02}, finding general expressions of the results of Eq. \eqref{fidelity} comes across challenges. For some special cases, benefit from the number expressions of the series expansions for $\lambda$, we find the results can be written as the forms which are listed in the the following.

Defining the function as
\begin{equation}\label{appendix03}
    f^{(m,n)}\equiv\frac{\lambda^{2n}}{N_{r,m,n}^{S}}\frac{m!n!(1+\lambda)}{2^{m+n+1}(1-\lambda)^{m+n}},
\end{equation}
we have

(i) For $m=0$ and $n\geq m$, the fidelity takes
\begin{equation}\label{appendix02}
    F=f^{(0,n)}.
\end{equation}
For resouces of the TMSVs, it is simpliy reduced to $F(\lambda)=f^{(0,0)}=(1+\lambda)/2$.

(ii) For $m=1$ and $n\geq m$, the fidelity takes
\begin{equation}\label{appendix04}
    F^{(1,n)}=\left[ \frac{2^2}{1}n-4n \lambda +(n+1)\lambda ^2\right]f^{(1,n)}.
\end{equation}

(iii) For $m=2$ and $n\geq m$, the fidelity takes
\begin{align}\label{appendix05}
    F^{(2,n)}=&\bigg[\frac{2^4}{2!}n(n-1)-\frac{2^4}{1}n(n-1)\lambda +4n(3n-1) \lambda ^2- 4n(n+1)\lambda ^3+ \frac{(n+1)(n+2) }{2!}\lambda ^4\bigg] f^{(2,n)}.
\end{align}

(iv) For $m=3$ and $n\geq m$, the fidelity takes
\begin{align}\label{appendix06}
    F^{(3,n)}=&\bigg[\frac{2^6}{3!}n(n-1)(n-3)-\frac{2^6}{2!}n(n-1)(n-3)\lambda+8n(n-1)(5n-7)\lambda^2-\frac{16}{3}n(n-1)(5n-1)\nonumber\\
    &+2n(n+1)(5n-2) -2n (n+1)(n+2)\lambda ^5+ \frac{(n+1)(n+2)(n+3)}{3!}\lambda ^6\bigg] f^{(2,n)}.
\end{align}

(v) For $m=4$ and $n\geq m$, the fidelity takes
\begin{align}\label{appendix07}
    F^{(4,n)}=&\bigg[\frac{2^8}{4!} n(n-1)(n-2)(n-3)-\frac{2^8}{3!} n(n-1)(n-2)(n-3)\lambda +\frac{32n(n-1)(n-2)(7n-17)}{3}\lambda ^2\nonumber\\
    &-\frac{32n(n-1)(n-2)(7n-9)}{3 }\lambda ^3+\frac{4}{3}n(n-1)\left(35n^2-55n+6\right)\lambda ^4-\frac{8}{3} n(n-1) (n+1) (7 n-2)\lambda ^5\nonumber\\
    &+ \frac{2n(n+1)(n+2)(7n-3)}{3}\lambda ^6-\frac{2}{3} n(n+1)(n+2)(n+3)\lambda ^7+\frac{(n+1)(n+2)(n+3)(n+4)\text{  }}{4!}\lambda ^8\bigg] f^{(4,n)}.
\end{align}

(vi) For $m=5$ and $n\geq m$, the fidelity takes
\begin{align}\label{appendix08}
    F^{(5,n)}=&\bigg[\frac{2^{10}}{5!}n(n-1)(n-2)(n-3)(n-4)- \frac{2^{10}}{4!}n(n-1)(n-2)(n-3)(n-4)\lambda \nonumber\\
    &+\frac{32}{3}n(n-1)(n-2)(n-3)(9n-31)\lambda ^2-\frac{128}{9}n(n-1)(n-2)(n-3)(9n-21)\lambda ^3\nonumber\\
    &+\frac{16}{3}n(n-1)(n-2)\left(62-77 n+21 n^2\right)\lambda ^4-\frac{16}{15} n(n-1) (n-2) \left(63n^2-91n+6\right)\lambda ^5\nonumber\\
    &+\frac{4}{3} n(n-1) (n+1) \left(21n^2-35n+6\right)\lambda ^6-\frac{8}{3} n(n+1)(n+2)(n-1)(3n-1)\lambda ^7\nonumber\\
    &+\frac{1}{6} n(n+1)(n+2)(n+3)(9n-4)\lambda ^8-\frac{1}{6}n(n+1)(n+2)(n+3)(n+4)\lambda ^9\nonumber\\
    &+ \frac{1}{5!}(n+1)(n+2)(n+3)(n+4)(n+5)\lambda ^{10}\bigg] f^{(5,n)}.
\end{align}

\end{document}